\documentstyle[prl,aps]{revtex}
\input epsf
\begin{document}

\draft

\title{Relaxing the Geodesic Rule in Defect
Formation Algorithms}


\author{Levon Pogosian and
Tanmay Vachaspati}
\address{
Physics Department, 
Case Western Reserve University, 
Cleveland OH 44106-7079.
}

%\date{\today}

\twocolumn[

\maketitle

%\tightenlines

\begin{abstract}
\widetext

In studying the formation of topological 
defects, it is conventional to assume the ``geodesic rule'' 
which is equivalent to minimizing gradients of the order parameter.
This assumption has been called into question in field-theoretic
studies of first order phase transitions and in the case of
local defects. We present a scheme for numerically investigating 
the formation of strings without assuming the geodesic rule. 
Our results show that the fraction of string in infinite
strings grows as we deviate from the geodesic rule.

\end{abstract}

%insert suggested PACS numbers in braces on next line

]

\narrowtext

Within all known theoretical frameworks, the early universe
must have been a close analog of condensed matter systems that
we now study in the laboratory. In particular, the early
universe must have undergone phase transitions during which
it is very conceivable that topological defects would have
been produced. The observation of any left-over defects today
would directly provide us with a window to the universe when
it was a mere fraction of a nanosecond old.

The study of topological defects has made long strides over
the last twenty years \cite{avps}. Starting with order-of-magnitude 
estimates, we are now in a position to compare theory with laboratory
experiments and astronomical observations. As our tools get
sharper, finer details of our understanding of the properties of 
defects will be tested.

The properties of defects at formation, {\it i.e.} at the phase 
transition, are an important ingredient in their cosmology. For
example, if the location of magnetic monopoles is very strongly
correlated with the location of antimonopoles, rapid annihilation
would ensue and the monopole over-abundance problem in cosmology
would be absent. Similarly, the presence of infinite strings in
cosmic string networks is expected to be very important in the 
cosmological role that they can play.

The properties of defects at formation have mostly been studied via 
lattice-based numerical simulations using the so-called Vachaspati-Vilenkin 
(VV) algorithm \cite{tvav}. To describe these efforts on a concrete footing, 
consider the global $U(1)$ model
\begin{equation}
L = |\partial _\mu \phi |^2 - {\lambda \over 4} (|\phi |^2 -\eta^2 )^2
\label{globalmodel}
\end{equation}
where $\phi = |\phi | e^{i\alpha}$ is a complex scalar field. Apart from
the discretization of space in the form of a specific lattice, the simulations 
also assume:
\begin{itemize}
\item Only the dynamics of the phase of $\phi$ is relevant. In other
words, $|\phi |$ is set to be $\eta$ everywhere.
\item The ``Geodesic Rule'': The gradients of $\alpha$ cost a lot of energy 
and are minimized.
\end{itemize}
Then the simulation proceeds by laying down phases $\alpha$ at every
lattice site, finding phase differences by minimizing gradients and
then evaluating
$$
\oint d\alpha
$$
around all plaquettes of the lattice. The value of this integral yields
the winding number and hence the number of strings passing through each
plaquette. Then the strings are joined together and in this way the string
network is constructed.

Very recently there has been a trend to study the formation of defects
and, in particular, strings, from a more field-theoretic 
viewpoint \cite{ams,ec,lp,af,ke,mh} and 
without adopting a lattice \cite{jbtkavtv,jb}. 
Both efforts are very desirable and have shed further light on the problem. 
For example, the study of the evolution of
fields during first order phase transitions (in two and three bubble
collisions) has shown that the degree of freedom associated with $|\phi |$ 
can be important in the formation of defects. Generally, oscillations of 
$|\phi |$ lead to an excess of strings over a simple counting of the 
winding of $\alpha$ as in the VV algorithm. Also, it is 
often found that the number of defects is in excess of what the 
geodesic rule would imply. So, it seems that both assumptions of the 
VV algorithm might need to be relaxed. 

Another situation where the geodesic rule has been called into question
is in the formation of gauge (local) strings \cite{rudazams}. 
In this case, the model in eq. (\ref{globalmodel}) 
is gauged. A consequence of this is that ${\bf \nabla} \alpha$ is no 
longer a gauge invariant quantity. For example, the
contribution to the energy density no longer comes 
from the ordinary gradient of $\alpha$ but from the covariant gradient: 
$$
({\bf \nabla} \alpha - e {\bf A} )^2
$$
where ${\bf A}$ is the gauge field. And this questions the 
geodesic rule that is based on the ordinary gradient of $\alpha$.

A significant advance has been made in Ref. \cite{tkav} where the
collision of bubbles of true vacuum is studied analytically. The end
result justifies the geodesic rule under certain circumstances. However,
in the analysis, the gauge field is not provided with any thermal properties 
of its own and is a passive player during the phase transition.
In general, we would expect that the gauge field 
would have its own dynamics consistent with the thermal nature of the system.
This could lead to the production of string in excess of what the geodesic
rule would imply.

\begin{figure}[tbp]
\caption{ \label{rhobeta} The filled circles show the string density
as a function of $\beta$ in the Monte Carlo simulation. The curve
shows the results of a direct calculation of the string density
where we evaluated the expectated number of strings passing through a
plaquette by summing over all phase distributions and values of
$n$ with appropriate probability factors.
}
\epsfxsize = \hsize \epsfbox{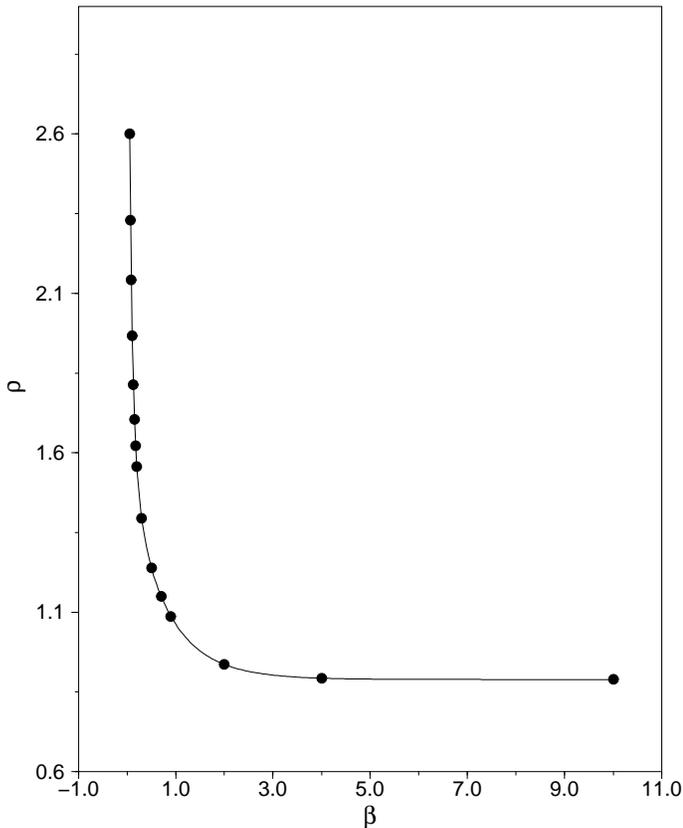}
\end{figure}

In this paper, we remedy the VV algorithm by relaxing the geodesic rule.
Instead of calculating the gradient in the phase $\alpha$ on a lattice
link by choosing the shortest path on the vacuum manifold, we choose the
path using a probability distribution. More explicitly, if the phase at
lattice site $A$ is $\alpha_A$ and that at lattice site $B$ is $\alpha_B$,
then the difference in phase between sites $B$ and $A$ is:
$$
\Delta \alpha = \alpha_B - \alpha_A + 2\pi n \equiv \delta \alpha +2\pi n
$$
where, $n$ is any integer. The geodesic rule was that $n$ should be chosen
so as to minimize $|\Delta \alpha |$. Here we take $n$
to be a random variable with the (Gaussian) probability distribution:
$$
P_n = \int_{n-0.5}^{n+0.5} dm ~
2 \sqrt{\pi \beta} e^{-\beta ( \delta \alpha + 2\pi m )^2} 
$$
where, $\beta \in (0,\infty )$ is a parameter.

The form of $P_n$ has two features which we think are desirable.
First, for large $\beta$, the algorithm reduces to the VV algorithm
with the geodesic rule. Secondly, the Gaussian form is reminiscent
of the Boltzmann suppression factor with $\beta$ being related to the
inverse temperature. On dimensional grounds, the connection of $\beta$
and the temperature $T$ is:
$$
\beta \sim {{\eta^2 \xi} \over T}
$$
where $\xi$ is the correlation length at the phase transition.

\begin{figure}[tbp]
\caption{ \label{frarho} The fraction of infinite string as a function
of the total string density. The 1$\sigma$-error bars are found by
running the simulation for each value of $\beta$ 100 times and then finding
the spread in the fraction of infinite strings. The spread in values of
the total string density is much smaller and has not been shown.
}
\epsfxsize = \hsize \epsfbox{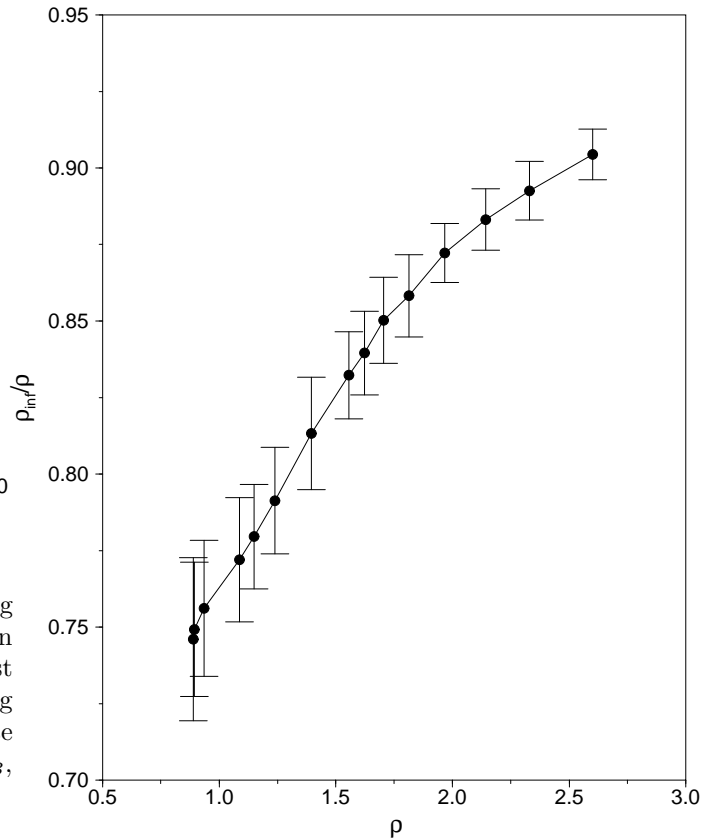}
\end{figure}

By changing the value of the parameter $\beta$, we can now study
the effects of relaxing the geodesic rule. This is equivalent to
allowing for the production of extra strings in bubble collisions,
say due to fluctuations of $|\phi |$, or, due to
gauge field fluctuations.

We have performed the numerical simulations in this way on a cubic
lattice together with the triangular discretization of phases described
in \cite{tvav}. Now it is possible for more than one string to pass 
through a plaquette in the lattice. This means that we can have a very high
density of strings. Also, in connecting the string network, at every
step there may be several choices to make. We make this choice randomly
with equal probability assigned to every possible connection.

The total string density in the Monte Carlo simulation has 
been plotted as a function of the parameter $\beta$ in Fig. 1. On this
figure we have also plotted the string density calculated by directly
evaluating the expected number of strings passing through each
plaquette. The two calculations are in agreement, giving us confidence
in the Monte Carlo simulations.

\begin{figure}[tbp]
\caption{\label{looplength} The loop length distribution for
$\beta =0.05$.
}
\epsfxsize = \hsize \epsfbox{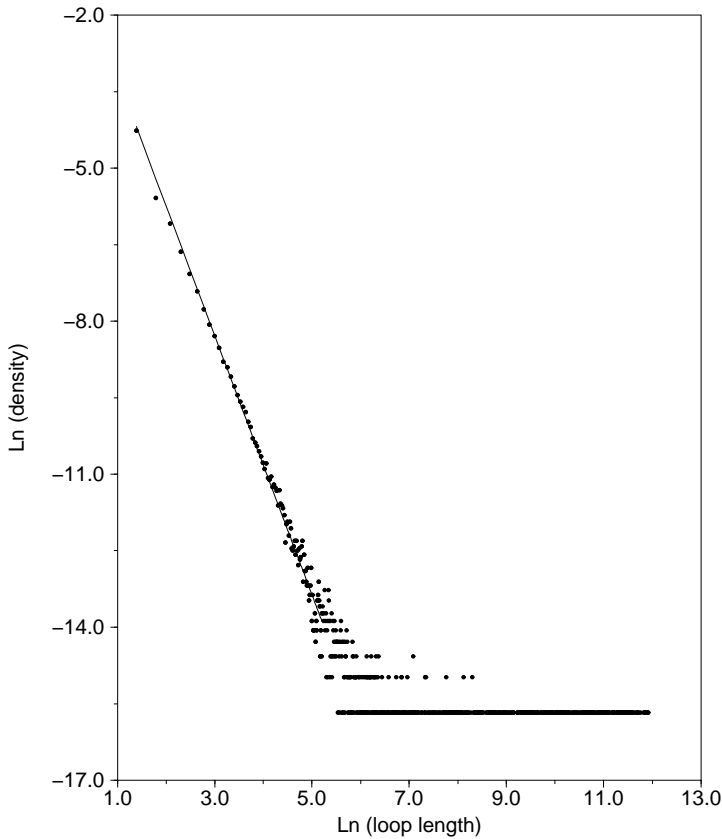}
\end{figure}

In Fig. 2, we show the string fraction in infinite string
($\rho_{inf} /\rho$) as a function of the total string
density $\rho$ where an infinite string is defined to be one that is
longer than the square of the lattice size. It is clear from the
graph that departures from the geodesic rule - smaller values
of $\beta$ - lead to an excess of infinite string fraction.

%In Fig. 2, we show the fraction of string density in infinite string, that
%is, $\rho_{inf} /\rho$, versus the parameter $\beta$. We have
%chosen to define an infinite string as one that is longer than the square
%of our lattice dimension. 

The graph in Fig. 2 can be fit by several different functional forms
and the data does not convincingly point to any fit. In principal it
should be possible to collect more data for higher values of the
string density (lower values of $\beta$) and then fit to a power law
in the limit $\rho_{inf} /\rho \rightarrow 1$. We have not been able
to find the corresponding exponent since data collection rapidly
becomes very expensive on computer resources as $\beta$ is reduced.

We have also plotted the number density of string loops as a function
of the length ($l$) of the loop. For all values of $\beta$, the result 
is the usual scale invariant distribution where the number density falls 
off as $l^{-2.5}$. In Fig. 3 we show this length distribution (for
$\beta =0.05$) which is well fitted by:
$$
dn = 0.5 {{dl} \over {l^{5/2}}} \ .
$$

To conclude,
our results show that the implementation of the geodesic rule in the VV
algorithm provides a lower bound on the fraction of infinite strings in
lattice simulations. It also shows that any fluctuations in the gauge
fields or magnitude of $\phi$ that produce more strings, are likely to
increase the fraction of infinite string.

{\it Acknowledgements:} TV was supported by the Department of Energy.

\end{document}